\documentstyle[epsf]{l-aa}

\begin{document}

\thesaurus{02(08.12.2; 10.07.2; 10.08.1; 12.04.1)}
\title{Zero-metallicity very low mass stars as halo dark matter}
\author{E.J. Kerins}
\institute{Observatoire de Strasbourg, 11 Rue de l'Universit\'e, 
F-67000 Strasbourg, France.}
\date{Received 21 October 1996 / Accepted 6 December 1996}
\offprints{{\tt kerins@wirtz.u-strasbg.fr}}
\maketitle

\newcommand{\sm}{\mbox{M}_{\sun}}
\newcommand{\tst}{\textstyle}
\newcommand{\be}{\begin{equation}}
\newcommand{\ee}{\end{equation}}

\begin{abstract}
Hubble Space Telescope (HST) limits on the amount of halo dark matter
(DM) in the form of very low-mass (VLM) stars close to the
hydrogen-burning limit have been interpreted as excluding such stars
as viable candidates. However, these limits assume that the stars are
smoothly distributed and have at least the metallicity of
Population~II stars, whilst some baryonic DM formation theories
predict that they may instead be clumped into globular-cluster
configurations and have close to zero metallicity. I re-analyse the
HST data employing the zero-metallicity VLM star models of Saumon et
al. (1994), which predict $V-I$ colours below the cuts of previous
analyses for stars below $0.2~\sm$. From the models I derive new
limits on the allowed halo fraction comprising VLM stars for both the
unclustered and clustered cases. In the unclustered regime I find a
95\% confidence upper limit on the allowed halo fraction of 1.4\%
inferred from 20 HST fields, comparable to limits derived by previous
studies for non-zero metallicity populations. In the cluster scenario
I show that clusters of mass $M$ and radius $R$ can satisfy both HST
and the recent MACHO gravitational microlensing results, which
indicate a lens halo fraction of 40\% for a standard halo model,
provided $R \la 1.2\,(M/10^4~\sm)^{0.74}$~pc. However, existing
dynamical limits restrict the allowed range to a tiny region
characterised by $M \sim 4\times 10^4~\sm$ and $R \sim
3$~pc. Furthermore, consistency between MACHO and HST demands a
present-day clustering efficiency of 92\% or better. Intriguingly
however, the cluster mass implied by these limits is theoretically
well motivated and the VLM star scenario may also help to provide an
explanation for the faint red `halo' light recently reported around
another galaxy.
\keywords{stars: low-mass, brown dwarfs -- globular clusters: general -- 
Galaxy: halo -- dark matter}
\end{abstract}

\section{Introduction} \label{s1}

Recent advances in attempts to detect or constrain the nature of dark
matter (DM) in the halo of our Galaxy have led to a variety of
constraints on both baryonic and non-baryonic candidates (Carr
\cite{carr94}; Jungman et al. \cite{jung96}).

The microlensing rate observed by the MACHO collaboration (Alcock et
al. \cite{alc97}) towards the Large Magellanic Cloud (LMC) indicates
that the likely halo DM fraction in compact form is around 40\%, and
comprises objects in the mass range $0.1 - 1~\sm$, though the inferred
fraction and mass range are sensitive to the assumed halo distribution
function. This therefore suggests that the halo may comprise a roughly
equal mixture of baryonic and non-baryonic matter, though there are
non-baryonic candidates, such as primordial black holes and shadow
matter, which can also explain the microlensing events. In any case,
the MACHO results imply that no {\em single\/} candidate, baryonic or
non-baryonic, can explain {\em all\/} of the halo DM unless the halo
distribution function departs significantly from the usual assumption
of an isothermal sphere.

If the DM responsible for the observed microlensing is baryonic then
the inferred mass range implicates either white-dwarf remnants of an
early generation of stars or very low-mass (VLM) stars close to the
hydrogen-burning limit. However, in our Galaxy the number density of
halo white dwarfs is strongly constrained by the present-day helium
and metal abundances of the interstellar medium (Carr et
al. \cite{carr84}; Ryu et al. \cite{ryu90}; Adams \& Laughlin
\cite{adam96}) and, assuming the white dwarfs are younger than 18~Gyr,
by number counts of high-velocity white dwarfs (Chabrier et
al. \cite{chab96}). Counts of high-redshift galaxies also appear to
rule out white dwarfs from contributing significantly to halo DM in
other galaxies (Charlot \& Silk \cite{char95}).

VLM stars are currently even less favoured since photometric surveys
appear to place even stronger limits on their contribution
(e.g. Gilmore \& Hewett \cite{gil83}; Richstone et al. \cite{rich92};
Bahcall et al. \cite{bah94}, hereafter BFGK; Graff \& Freese \cite{graf96},
hereafter GF; Flynn et al. \cite{fly96}). The latest MACHO results
also appear to argue against most of the DM comprising sub
hydrogen-burning brown dwarfs unless the DM is distributed more like a
maximal disc, in which case any halo must have a very low total mass
interior to the LMC in order to remain consistent with local
surface-density and rotation-curve constraints. Within the context of
standard halo models therefore, there is great difficulty in
explaining the high halo fraction detected by MACHO by invoking a
single baryonic candidate.

In this paper I re-examine the VLM star hypothesis for the case
where the stars have zero-metallicity, as may be the case if the
halo is formed from primordial matter. In this case their colours
will be different than assumed by previous studies, so constraints
from these studies need not necessarily apply to zero-metallicity
stars. I also examine the possibility of DM clustering, and
how that effects number counts and microlensing statistics. Such
clustering is predicted by some DM formation theories.

The plan of the paper is as follows: Sect.~\ref{s2} calculates the
space density constraints for a smoothly distributed population of
zero-metallicity VLM stars imposed by number counts from 20 of the 22
HST fields obtained by BFGK. These calculations adopt the latest
numerical predictions for zero-metallicity VLM star photometry. It
will be shown that the limits on such a population are comparable to
those for populations of non-zero metallicity, with the allowed halo
fraction being at most 1.4\%.  Section~\ref{s3} extends the analysis
to include the effect of clustering and, using a combination of the
HST number-count limits, MACHO observations and dynamical constraints
on the permitted mass and radius of clusters, places upper limits on
the cluster space density. It will be shown that whilst dynamical
constraints strongly limit the allowed mass and radius of clusters,
they nonetheless permit a cluster DM fraction which can simultaneously
explain the MACHO and HST observations, though both the required mass
$M$ and radius $R$ need to be finely tuned ($M \sim 4 \times
10^4~\sm$, $R \sim 3$~pc). I further show that the microlensing
signatures of these clusters are essentially indistinguishable from
the unclustered scenario, and in particular are not expected to modify
the inferred halo fraction. Finally in this section, I use the halo DM
fraction inferred by MACHO to compute the efficiency with which VLM
stars must be clustered in order to remain compatible with HST
observations. The required efficiency, better than 92\% at the present
day, places the strongest demand on the scenario. The main findings of
the paper are discussed in Sect.~\ref{s4}.

\section{Limits on zero-metallicity VLM stars} \label{s2}

Saumon et al. (\cite{sau94}) have computed a grid of fully non-gray
atmospheric models for zero-metallicity ($Z = 0$) VLM stars and brown
dwarfs in the mass range $0.01 - 0.2~\sm$. Their calculations take
account of all the major sources of opacity for a pure H+He mixture
including H$_2$, H and H$^-$, as well as other sources tabulated by
Lenzuni et al. (\cite{len91}). The resulting spectral energy
distributions are found to deviate significantly from blackbody below
an effective temperature of 4\,000~K. Saumon et al. also present
predictions for the photometry of VLM stars ($0.092 - 0.2~\sm$) from
$M$ through to $V$ bands which show that these stars would appear
significantly bluer than stars with metallicity comparable
to that measured for the Galactic spheroid ($Z \sim 0.01~Z_{\sun}$) or
disc ($Z \sim Z_{\sun}$) populations.

In particular, Saumon et al. find that the expected $V-I$ colour for
zero-metallicity VLM stars ranges between 1.27 for 0.2-$\sm$ stars to
1.57 at the hydrogen-burning limit ($0.092~\sm$). This is somewhat
below the threshold $V-I$ values assumed by BFGK, GF and Flynn et
al. (\cite{fly96}) in their analyses, whose inferences are based on
the number counts of stars with $V - I \gse 2$. Whilst one can not say
for sure whether any dark halo population has a metallicity
substantially less than that of the spheroid, there is
clearly a need to re-examine the HST data using the zero-metallicity
predictions of Saumon et al. in order to close up this last remaining
`loophole'.

The positions of the 22 HST fields are tabulated in Gould et
al. (\cite{gou96}), along with the corresponding minimum and maximum
$I$-band magnitudes for each field. (All selections were performed in
the $I$ band.) Note that two locations ($l = 82\degr$, $b = -19\degr$
and $l = 134\degr$, $b = -65\degr$) each have 2 fields in very close
proximity.

The limiting maximum $I$-band magnitude, which determines the maximum
distance out to which a VLM star will be seen, ranges from 22.56 to
24.40 and is determined for each field according to the ability to
discriminate clearly between stellar and extended sources. The minimum
$I$-band magnitude ranges from 17.05 to 19.45 and sets the minimum
distance at which the stars can be satisfactorily imaged. Objects with
magnitudes below this limit produce saturated images. It should be
noted that BFGK calculate the $I$-band magnitude limits from the HST
$I'$ band (F814W filter) assuming the stellar spectral energy
distributions listed by Gunn \& Stryker (\cite{gun83}). They therefore
do not {\em strictly\/} apply to $Z = 0$ VLM stars, though any
differences in calibration will be small and are therefore neglected
here.

In computing the minimum and maximum observable distances for each VLM
star mass, and for each field, I follow Gould et al. (\cite{gou96}) in
converting the $B$-band extinction values determined by Burstein \&
Heiles (\cite{bur82}) to $I$-band reddenings. The extinction is
assumed to be confined to a disc of thickness $|z| = 100$~pc. Within
this disc the extinction per unit distance is taken to be uniform.

The halo mass within the volume defined by the minimum and maximum
distances $d_{\rm min}$, $d_{\rm max}$, and the solid angle per
field $\Omega = 3.7\times 10^{-7}$~sr, is calculated assuming a 
spherically-symmetric softened isothermal halo density distribution of
the form
   \be
      \rho(x,l,b) = \rho_0 \left( \frac{a^2 + R_0^2}{x^2-2xR_0 \cos l
      \cos b +a^2 + R_0^2} \right), \label{e1}
   \ee
where $x$ is the distance measured along the observer's line of sight,
$l$ and $b$ are Galactic coordinates, $R_0 = 8$~kpc is the Sun's
Galactocentric distance, $\rho_0 = 0.01~\sm~\mbox{pc}^{-3}$ is taken
to be the local DM density normalisation and $a = 5$~kpc is the
assumed halo core radius. For the small solid angles considered here,
this gives an integrated halo mass between $d_{\rm min}$ and $d_{\rm
max}$ for field $i$ of
   \be
      M_{{\rm halo},i} \simeq \frac{\pi \Omega}{4} \int_{d_{{\rm
      min},i}}^{d_{{\rm max},i}} x^2 \rho(x,l_i,b_i) \, {\rm d}x.
      \label{e2}
   \ee
The integral can be performed analytically though the resulting
expression is long.

Note that the integral limits in Eq.~\ref{e2} are implicit functions
of the VLM star mass $m$. If one assumes that VLM stars have masses
$m_{\rm *} \ge m$ then a lower limit on the expected number of
detectable VLM stars is
   \begin{eqnarray}
      N_{\rm VLM}(\ge m) & \ge & \frac{1}{m} \sum_{i=1}^n
         M_{{\rm halo},i}(m) \nonumber \\
      & \gse & \frac{\pi \Omega}{4m} \sum_{i=1}^n \left[ \int_{d_{{\rm
      min},i}(m)}^{d_{{\rm max},i}(m)} x^2 \rho(x,l_i,b_i) \, {\rm d}x
      \right] \label{e3}
   \end{eqnarray}
for $n$ independent fields. The direction of the inequality reflects
the fact that the dependency of $M_{{\rm halo},i}$ on $m$ is steeper
than the first power of $m$. Fields~1 and 19, using the order in which
they are listed in Tab.~1 of Gould et al. (\cite{gou96}), are
discarded in this analysis because of their close proximity to fields
2 and 20, respectively. Whilst fields~1 and 2 do not actually overlap,
field~1 is nonetheless excluded here to provide consistency with the
cluster analysis of Sect.~\ref{s3.2}, where statistical independency of
neighbouring fields is an important criteria. $N_{\rm VLM}$ is
therefore summed over $n = 20$ rather than 22 fields.

The median value for $d_{\rm min}$ is found to range from 330~pc for
0.092-$\sm$ stars up to 1~kpc for 0.2-$\sm$ stars. For $d_{\rm max}$
the median values are 3.3~kpc and 10.1~kpc for 0.092-$\sm$ and
0.2-$\sm$ stars, respectively.

Applying Eq.~\ref{e3} to VLM stars at the hydrogen-burning limit mass
of $0.092~\sm$, one finds an expectation number of stars in the HST
fields of 6\,310, and for 0.2-$\sm$ objects the expectation is nearly
an order of magnitude larger at 60\,100 stars. These numbers take
account of the fact that data from one-third of field 4 (i.e. data
from one of the three detector chips) had to be discarded by Gould et
al. (\cite{gou96}) due to problems with receiving the data from HST,
and that as much as 2\% of each of the fields was discarded due to
emission from background galaxies.  The number of stars detected in
the 20 fields with $V-I$ values in the range 1.2 to 1.7, spanning the
range predicted by Saumon et al. (\cite{sau94}), is only 75. The
95\% confidence level (CL) upper limit on the average, for a
realisation of 75 stars, is 91. Therefore, even if one assumes that
all of the stars detected by HST are halo VLM stars right on the
hydrogen-burning limit, their contribution to the halo DM can be no
more than 1.4\% at the 95\%~CL, and the limit is correspondingly
stronger than this for more massive objects. In fact it is likely that
a significant fraction of these objects may belong to the disc or
spheroid.

The limit of 1.4\% is stronger than that inferred by BFGK for
solar-metallicity VLM stars from their analysis of one of the HST
fields. Their results translate to a 95\%~CL upper limit of less than
4\% for stars with $V-I > 3$, adopting $\rho_0 =
0.01~\sm~\mbox{pc}^{-3}$.  The 95\% upper limit inferred by GF for
low-metallicity stars with $2 < V-I < 3$ corresponds to a halo
fraction of less than 0.9\% for the same field. However, both of these
studies exclude from their analyses stars with $V-I < 2$, so the
limits presented here, which are derived from 20 fields, constitute a
completely independent check on previous results. The conclusion is
that, regardless of their metallicity, VLM stars do not contribute
significantly to the halo DM, at least under the assumption that they
are smoothly distributed in the halo.

\section{VLM star clusters} \label{s3}

\subsection{Rationale} \label{s3.1}

The rationale for invoking clustered DM comes from a number of
considerations.  Firstly, visible globular clusters are observed out
to great distances from the Galactic centre suggesting they may have
an important role in halo formation, though they are identified with
the spheroid population. Secondly, some of the most promising theories
for the formation of large amounts of compact halo baryonic DM (Ashman
\cite{ash90}; De~Paolis et al. \cite{dep95}) predict that the DM
should be clustered into dark globular clusters with a typical mass of
around $10^4~\sm$ (Ashman \cite{ash90}). These theories predict that
the clusters should comprise either brown dwarfs or VLM stars, due to
high gas pressures suppressing the minimum fragmentation mass. They
also explain the typical mass and spatial distribution of the visible
cluster population since the theories are extensions of the Fall-Rees
theory of globular-cluster formation (Fall \& Rees \cite{fall85}). The
baryon Jeans mass at the cosmological epoch of recombination
is $M_{\rm BJ} = 1.3 \times 10^6 \, \Omega_{\rm B}
\Omega_0^{1/2} h^{-1}~\sm$, which is also close to $10^4~\sm$ for a
cosmological density $\Omega_0$ of order unity and a baryon density
$\Omega_{\rm B}$ satisfying cosmological nucleosynthetic constraints:
$0.01 < \Omega_{\rm B} h^2 < 0.024$, where $h$ is the Hubble constant
in units of 100~km~s$^{-1}$~Mpc$^{-1}$ (Krauss \cite{kra95}). Lastly,
limits on the space density of clusters provide {\em firm\/} upper
limits on the allowed space density of VLM stars, provided one assumes
that the VLM star distribution follows the halo density on the
large-scale average (i.e. scales larger than the typical cluster
separation).

The effect that clustering can have on number-count statistics is well
illustrated by considering the expectation distances both to the
nearest unclustered VLM star and to the nearest cluster.
Let us assume that VLM stars contribute some fraction $f_{\rm h}$ to
the halo DM and that their density is everywhere $f_{\rm h}\rho$,
where $\rho$ is given by Eq.~\ref{e1}. Let us also take the fraction
of VLM stars residing in clusters to be everywhere $f_{\rm c}$, so
that the fraction of unclustered VLM stars is correspondingly $1 -
f_{\rm c}$.  Then, for VLM stars with mass $m$, the local unclustered
number density is
   \begin{eqnarray}
      n_{{\rm u},0} & = & \frac{f_{\rm h} (1-f_{\rm c}) \rho_0}{m} 
         \nonumber \\
      & = & 0.1 \, f_{\rm h}(1-f_{\rm c}) \left( \frac{m}{0.1~\sm}
      \right) ^{-1} ~\mbox{pc}^{-3}, \label{e4}
   \end{eqnarray}
whilst the local number density of VLM star clusters of mass $M$ is
   \begin{eqnarray}
      n_{{\rm c},0} & = & \frac{f_{\rm h}f_{\rm c} \rho_0}{M} 
         \nonumber \\
      & = & 10^{-6} \, f_{\rm h} f_{\rm c} \left( \frac{M}{10^4~\sm}
      \right) ^{-1} ~\mbox{pc}^{-3}. \label{e5}
   \end{eqnarray}
The expectation value for the distance $d$ to the nearest object
appearing within a survey of solid angle $\Omega$, assuming a local
number density $n_0$, is (c.f. Kerins \& Carr \cite{ker94})
   \begin{eqnarray}
      \langle d \rangle & \equiv & \Omega n_0 \int_0^\infty x^3
         \exp (-\Omega n_0 x^3/3) \, {\rm d}x \nonumber \\
      & = & ( 3/\Omega)^{1/3} \Gamma(4/3) n_0^{-1/3}, \label{e6}
   \end{eqnarray}
where $\Gamma(4/3) = 0.893$ is the gamma function. So, using
Eq.~\ref{e4} and Eq.~\ref{e5}, the expectation distances to the
nearest unclustered VLM star and star cluster are, respectively,
   \begin{eqnarray}
      \langle d_{\rm u} \rangle & = & 0.14 \, f_{\rm h}^{-1/3}
         (1 - f_{\rm c})^{-1/3} \nonumber \\
         & & \times \left( \frac{\Omega}{7.4 \times
         10^{-6}~\mbox{sr}} \right)^{-1/3} \left( \frac{m}{0.1~\sm}
         \right)^{1/3}~\mbox{kpc}  \label{e7} \\
      \langle d_{\rm c} \rangle & = & 6.6 \, f_{\rm h}^{-1/3}
         f_{\rm c}^{-1/3} \nonumber \\
         & & \times \left( \frac{\Omega}{7.4 \times
         10^{-6}~\mbox{sr}} \right)^{-1/3} \left( \frac{M}{10^4~\sm}
         \right)^{1/3}~\mbox{kpc}, \label{e8}
   \end{eqnarray}
where $\Omega$ is normalised to a solid angle 20 times that of a
single HST field.

The large value for $\langle d_{\rm c} \rangle$ indicates the need to
consider the variation of $\rho$ along the line of sight in order to
obtain a more accurate answer, though the estimate of Eq.~\ref{e8} is
sufficient to show that one might not see {\em any\/} clusters at all
within the HST fields, even if they comprise all the halo
DM. Additionally, the expected fluctuation in source counts from field
to field will be governed by the number density of clusters and not
the spatially-averaged density of the stars themselves. This problem
is addressed in the next subsection where I calculate 95\%~CL lower
limits on the expected cluster number density, which in turn provide
95\%~CL upper limits on the contribution of VLM stars to the halo DM.

Another important consideration is whether one expects HST to be able
to resolve individual stars within clusters. Since BFGK filter their
observations to search for only stellar-like sources it is conceivable
that a significant fraction of the total number of possible VLM star
sources have been discarded because they reside in the cores of
unresolvable clusters. Since the number of field galaxies far exceeds
the number of stars at the typical limiting magnitudes of the HST
fields, it may prove extremely difficult to analyse these fields for
the presence of unresolved cluster cores.

Using Eq.~\ref{e8} as a lower limit on the cluster distance one finds
a typical angular separation between neighbouring stars (as projected
along the line of sight) for a cluster with radius $R$ of
   \begin{eqnarray}
      \theta & \la & 0.5 \, \langle d_{\rm c} \rangle^{-1} 
      \left( \frac{M}{\pi R^2 m} \right)^{-1/2} \nonumber \\
      & \la & 0.9 \, f_{\rm h}^{1/3} f_{\rm c}^{1/3} \left(
         \frac{\Omega}{7.4 \times 10^{-6}~\mbox{sr}} \right)^{1/3}
         \left( \frac{M}{10^4~\sm} \right)^{-5/6} \nonumber \\
         & & \quad \quad \quad \quad \quad \times \left( \frac{R}{10~\mbox{pc}}
         \right) \left( \frac{m}{0.1~\sm} \right)^{1/2}~\mbox{arcsec},
         \label{e9}
   \end{eqnarray}
compared to a HST pixel resolution of 0.1~arcsec. Since one expects
clusters to be centrally concentrated one anticipates
Eq.~\ref{e9} overestimating the angular separation of stars near the
core and underestimating the separation near the cluster
edge. Therefore it seems that a significant fraction of cluster
sources may appear unresolved.  This fraction will be assessed more
rigorously in Sect.~\ref{s3.3}.

For the cluster scenario to be viable it must be consistent with
existing dynamical constraints on such objects. These constraints,
which are reviewed in Sect.~\ref{s3.4}, provide upper and lower bounds
to both the permitted cluster mass and radius. The maximum
contribution of such clusters to the halo is therefore also
potentially bounded by these constraints.

Finally, in Sect.~\ref{s3.5}, I show that this scenario is compatible
with MACHO microlensing observations, and I use the MACHO limit on the
contribution of VLM stars to compute the allowed ratio of
clustered to unclustered stars. Potentially, this ratio provides the
strongest constraint on the cluster scenario.

\subsection{Space density of VLM clusters} \label{s3.2}

In Sect.~\ref{s2} limits were placed on the space density of unclustered
VLM stars from HST number counts, assuming the stellar distribution
obeys Poisson statistics. However, since clusters are extended
objects, one can not directly apply the same assumption to them, since
one may have a cluster which is only partially within the field of
view (in fact this will be the case generally for the HST fields due
to their small solid angle).

However one can use Poisson statistics to assess the likely number of
cluster {\em centres\/} within the field of view. One can therefore
extend this idea by imagining a volume centred on the actual field
volume, but whose radius of cross section along the length is always
larger by an amount $R$ (the cluster radius) and whose length is also
larger by an amount $R$ at either end (i.e. by an amount $2R$
overall). Limits on the number of cluster centres appearing within
this larger volume necessarily correspond to limits on the number of
clusters which appear {\em either wholly or partially\/} within the
actual survey field volume.

The description above applies to a cylindrical survey volume (i.e. a
circular field of view). The size of the larger volume and
number-count limits derived from it are sensitive to the geometry of
the field volume on which it is centred (i.e. the shape of the field of
view), which for the HST is `L'-shaped rather than
circular. However, for most of the cluster radii considered in this
paper their angular sizes are always much larger than the angular size
of the field of view itself, so the sensitivity to the shape of the
field of view is generally very small, and so for simplicity I assume
a cylindrical field volume geometry throughout.

Another important consideration when placing limits on cluster number
densities is the statistical dependency of neighbouring fields.
In principle, sufficiently large clusters could appear in more than one
field and so any limits would need to take such correlations into
account. Even if one is placing limits based on the {\em absence\/} of
clusters within the observed fields one still needs to consider this
problem. One can see why by considering two neighbouring fields which
are separated by less than a cluster diameter. The prescription above
would oversample the actual volume of space required to place limits
on the cluster density (i.e. there would be a volume of space which
would be common to the periphery of both fields and would therefore be
counted twice), resulting in an overestimate of the expected number of
clusters. As mentioned in Sect.~\ref{s2}, fields~1 and 2, whilst not
overlapping, are nonetheless too close for most of the cluster radii
considered here and so field~1, which samples a smaller amount of halo
mass than field~2 (using Eq.~\ref{e2}), is discarded.  Similarly,
field~19, which partially overlaps with field~20, is also discarded
leaving $n = 20$ fields.  For these remaining fields the spatial
separation between nearest-neighbours ranges from $150 - 960$~pc
for clusters comprising 0.2-$\sm$ VLM stars (with a median separation
of 340~pc), and from $50 - 310$~pc for clusters comprising 0.092-$\sm$
stars (with a median value of 110~pc). The statistical dependency of
neighbouring fields therefore only becomes important for the very
largest cluster radii considered in this section. I therefore ignore
such correlations and assume the fields to be statistically
independent.

A lower limit on the expectation number of clusters with mass
$M$ and radius $R$, comprising VLM stars of mass $m_{\rm *} \ge m$,
which either wholly or partially appear within the $n = 20$ HST fields
is (c.f. Eq.~\ref{e3})
   \be
      N_{\rm c}(M,R,\ge m) \ge \frac{f_{\rm c}}{M} \sum_{i = 1}^n M'_{{\rm
      halo},i} (R,m), \label{e10}
   \ee
where
   \begin{eqnarray}
      M'_{{\rm halo},i} & \simeq & \pi \int_{d_{{\rm
      min},i}(m)-R}^{d_{{\rm max},i}(m)+R} \left( \frac{\Omega^{1/2}x}{2}
      + R \right)^2 \rho(x,l_i,b_i) \, {\rm d}x. \nonumber \\
      & & \label{e11}
   \end{eqnarray}
Note that for $R = 0$, $M'_{{\rm halo},i} = M_{{\rm halo},i}$ in
Eq.~\ref{e2}, as required.

If one assumes that the HST fields contain no cluster stars at all
then the 95\%~CL upper limit on the allowed halo fraction
is $f_{\rm h} = 3\, f_{\rm c} N_{\rm c}^{-1}$; 3 being the 95\%~CL
upper limit on the true average number of clusters when none are seen,
assuming statistical independency of neighbouring fields and Poisson
statistics. Note from Eq.~\ref{e10} that the upper limit $f_{\rm h}$
is independent of $f_{\rm c}$ since it represents a limit on the {\em
absolute\/} number of clusters. If $f_{\rm c}$ is significantly less
than unity stronger constraints on the VLM density will come from the
limits obtained in Sect.~\ref{s2} for unclustered VLM stars. This
issue is considered in more detail in Sect.~\ref{s3.5} where I
consider the efficiency with which VLM stars need to be clustered.

\begin{figure}
\epsfxsize=8.5cm
\epsfbox{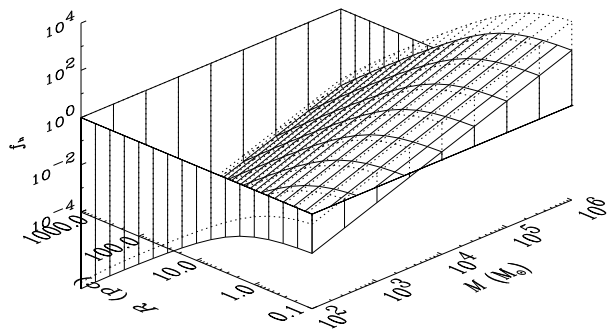}
%\picplace{8.0cm}
\caption[]{95\%~CL upper limits on the allowed halo fraction $f_{\rm
h}$, as a function of cluster mass $M$ and radius $R$, assuming a
clustering fraction $f_{\rm c} = 1$. The solid grid shows the limits
for clusters comprising VLM stars of mass $m = 0.2~\sm$ and the dotted
grid for stars at the hydrogen-burning limit, $m = 0.092~\sm$. $f_{\rm
h}$ is normalised to a local halo DM density $\rho_0 =
0.01~\sm~\mbox{pc}^{-3}$, and $f_{\rm h} = 1$ is demarcated by the
plane on which the skirting around the perimeter of the grid is
centred. The region of the grid at low cluster masses and large radii,
which dips below this plane, is constrained by HST number counts,
whereas the region rising above it denotes cluster parameters which
can provide all of the halo DM whilst remaining compatible with HST
observations.}
\label{f2}
\end{figure}

The limits on $f_{\rm h}$ for a range of cluster masses and radii are
shown in Fig.~\ref{f2}, assuming $f_{\rm c} = 1$. The solid grid depicts
the limits assuming the clusters comprise VLM stars of mass $m =
0.2~\sm$, where as the dotted line shows the limits for clusters
comprising hydrogen-burning limit stars ($m = 0.092~\sm$). The choice
of cluster parameters is guided by the dynamical considerations
discussed in Sect.~\ref{s3.4}. The plane at $f_{\rm h} = 1$ is shown
to guide the eye in seeing which part of the parameter space is
constrained and which is not. The most strongly constrained region
occurs for low cluster masses and large radii, though combinations of
cluster parameters for which $M < 4\pi \overline{\rho} R^3/3$ (where
$\overline{\rho}$ is the average halo density within the survey
volumes) are unphysical since they denote clusters whose average
density is less than that of the local halo background. (Such
`clusters' might instead best be regarded as voids.) For interesting
(i.e. significant) values of $f_{\rm c}$ and $f_{\rm h}$ this
unphysical region includes the case where the average angular
separation between neighbouring stars, given by Eq.~\ref{e9}, exceeds
the HST field of view. Whilst HST provides strong limits on clusters
of low mass and large radius, it is clear from the figure that a
significant fraction of the parameter space is not constrained by HST
observations.

It is apparent from Fig.~\ref{f2} that the relationship between
$f_{\rm h}$ and $M$ is linear, which is simply a consequence of the
fact that $f_{\rm h} \propto N_{\rm c}^{-1} \propto M$ from
Eq.~\ref{e10}. Whilst the relationship between $f_{\rm h}$ and $R$ is
not a simple power law one can crudely approximate the condition for
which $f_{\rm h}$ can exceed 1 as
   \be
      R \la \alpha \, \left( \frac{M}{10^4~\sm} \right) ^{\beta},
      \label{e12}
   \ee
where for 0.2-$\sm$ stars $\alpha \simeq 0.2$~pc and $\beta \simeq
1.05$, and for 0.092-$\sm$ stars $\alpha \simeq 1.3$~pc and $\beta
\simeq 0.75$. If instead of providing all of the halo DM one merely
requires that clusters provide at least 40\% of the DM, the halo
fraction indicated by the MACHO experiment for a local halo
normalisation $\rho_0 = 0.01~\sm~\mbox{pc}^{-3}$ (Alcock et
al. \cite{alc97}), then one requires $\alpha \simeq 0.4$~pc and $\beta
\simeq 0.97$ for 0.2-$\sm$ stars, or $\alpha \simeq 2.0$~pc and $\beta
\simeq 0.73$ for hydrogen-burning limit stars. These inequalities
should be compared to the requirement noted above that clusters must
represent local density {\em enhancements\/} rather than decrements,
which requires
   \be
      R < 62 \, \left( \frac{M}{10^4~\sm} \right) ^{1/3} \left(
      \frac{\overline{\rho}}{0.01~\sm~\mbox{pc}^{-3}} \right)^{-1/3}
      ~\mbox{pc}. \label{e13}
   \ee
The limit should also be compared to the dynamical constraints to be
discussed in Sect.~\ref{s3.4}.

It is important to note that these limits are fairly insensitive to
the actual number of VLM stars observed in the HST fields, provided
the observed number is relatively small (e.g. $N_{\rm VLM} \la
100$). This is because if even just a portion of one cluster enters
one of the fields of view one would generally expect HST to detect
hundreds or even thousands of stars in that field due to the high
cluster surface density. For the same reason these limits are also
relatively insensitive to the surface-density profile of the cluster,
at least within the range of reasonable surface-density profiles
(i.e. similar to those inferred from the surface-brightness profiles
of visible clusters).

\subsection{The question of resolvability} \label{s3.3}

The previous subsection dealt with limits imposed from the absence of
clusters within the HST fields. However, compact clusters can escape
detection even if they do appear in the fields, provided their surface
densities are sufficiently high as not to allow them to be resolved,
or provided the total surface density through all clusters along the
line of sight is sufficiently large. The rigorous point-source
selection criteria of BFGK means that any unresolved portions of
clusters are discarded, in which case only some fraction $f_{\rm
res} \leq 1$ is detectable. Thus the {\em effective\/} number of
detectable clusters is $N_{\rm c,eff} \equiv \langle f_{\rm res}
\rangle N_{\rm c} \leq N_{\rm c}$, which implies that the effective
allowed halo fraction is $f_{\rm h,eff} \propto \langle f_{\rm res}
\rangle^{-1} N_{\rm c}^{-1} \geq f_{\rm h}$, where $\langle \ldots
\rangle$ denotes averaging over fields. Thus if $f_{\rm res} = 0$ for
a particular set of cluster parameters $(M,R)$ then HST point-source
counts do not place any limits on the allowed halo fraction in such
clusters.

The effect on resolvability due to several clusters aligned along the
line of sight can be estimated from the ratio of the average cluster
surface density to that between $d_{\rm min}$ and $d_{\rm max}$:
   \be
      \gamma \equiv f_{\rm c}f_{\rm h} \frac{\pi R^2}{M} \int_{d_{\rm min}}
      ^{d_{\rm max}} \rho(x,l,b) \, {\rm d}x. \label{e13.1}
   \ee
Assuming the distribution of clusters on the sky is
Poissonian, their sky-covering factor is $1-\exp(-\gamma)$. Therefore the
surface density measured along a line of sight through a cluster will
be enhanced over that expected for the single cluster alone by a
factor which is on average
   \be
      \epsilon_{\sigma} = \gamma [1-\exp(-\gamma)]^{-1}. \label{e13.2}
   \ee
The median line-of-sight surface density between $d_{\rm min}$ and 
$d_{\rm max}$ for the 20 HST fields is $25.2~\sm~\mbox{pc}^{-2}$
for 0.092-$\sm$ stars and $64.6~\sm~\mbox{pc}^{-2}$ for 0.2-$\sm$
stars, corresponding to values for $\epsilon_{\sigma}$ of 1.45
and 2.34, respectively, assuming $10^4$-$\sm$ clusters with
a 10-pc radius and $f_{\rm c} = f_{\rm h} = 1$.

The resolvable fraction of an individual cluster is sensitive to its
assumed surface-density profile, as well as its mass and radius and
the distance at which it is expected to be observed. I assume here
that the cluster surface density follows the surface-brightness
profile of many observed globular clusters, which are well
described by the King (\cite{king62}) surface-brightness law. Thus the
surface number density $\sigma$ as a function of cluster-centric
radius $r$ is assumed to be given by
   \be
      \sigma = \left\{ \begin{array}{ll}
         \frac{\tst \sigma_0}{\tst 1 + (r/r_{\rm c})^2} & (r \leq R) \\
         0                                    & (r > R)
      \end{array} \right., \label{e14}
   \ee
where $\sigma_0 \equiv \sigma(r = 0)$ denotes the central surface
density and $r_{\rm c}$ is the projected cluster core radius. Since
the integrated mass of the cluster must be $M$ the central surface
density is
   \be
      \sigma_0 = \frac{(M/m)}{\alpha^2 \pi R^2 \ln (1+ \alpha^{-2})},
      \label{e15}
   \ee
where $\alpha \equiv r_{\rm c}/R$. For the purpose of simplification I
adopt $\alpha = 0.1$ for all clusters, a value which is typical for
visible clusters, though in reality there is a large dispersion about
this value.

Clusters appear resolved in the HST fields provided $\sigma_0$
is less than
   \begin{eqnarray}
      \sigma_{\rm res} & = & (2 x \theta_{\rm res})^{-2} \nonumber \\
          & = & 4.3 \times 10^{4} \left( \frac{x}{5~\mbox{kpc}}
      \right)^{-2}~\mbox{pc}^{-2}, \label{e16}
   \end{eqnarray}
where $\theta_{\rm res} = 0.1$~arcsec for the HST. If $\sigma_0 >
\sigma_{\rm res}$ then one expects that at least some of the cluster
is unresolvable. From Eq.~\ref{e14} to \ref{e16}, the mass
fraction of a cluster at distance $x$ which can be resolved is
   \begin{eqnarray}
      f_{\rm res}(x) & = & 1 - \frac{m}{M} \int_{0}
      ^{\min[r(\sigma = \sigma_{\rm res}/\epsilon_{\sigma}),R]} 2\pi r
      \sigma(r)\, {\rm d}r \nonumber \\
      & = & 1 - \min \left\{ 1, \frac{\ln [\epsilon_{\sigma} \sigma_0/
      \sigma_{\rm  res}(x)]}{\ln (1 + \alpha^{-2})} \right\},
      \label{e17}
   \end{eqnarray}
if $\epsilon_{\sigma} \sigma_0 > \sigma_{\rm res}$ or unity otherwise.
Note the factor $\epsilon_{\sigma}$ from Eq.~\ref{e13.2}, which enters
in the upper integral limit. This is because, as argued above, one
expects the observed line-of-sight density to be $\epsilon_{\sigma}
\sigma$, rather than just $\sigma$.

Thus, for HST field $i$, the expectation value for the resolvable
fraction of clusters is
   \be
      \langle f_{{\rm res},i} \rangle \simeq \frac{\int_{d_{{\rm min},i}}
      ^{d_{{\rm max},i}} x^2 \rho(x) f_{\rm res}(x) \, {\rm d}x}{\int
      _{d_{{\rm min},i}}^{d_{{\rm max},i}} x^2 \rho(x) \, {\rm d}x},
      \label{e18}
   \ee
where $\rho$ is given by Eq.~\ref{e1}. Hence, the {\em effective\/}
number of detectable clusters within the $n = 20$ fields is
given by Eq.~\ref{e10} with $M'_{{\rm halo},i} \rightarrow \langle 
f_{{\rm res},i} \rangle M'_{{\rm halo},i}$.

\begin{figure}
\epsfxsize=8.5cm
\epsfbox{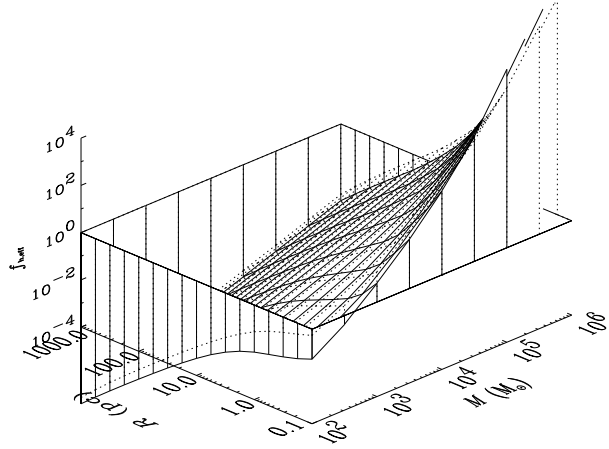}
%\picplace{8.0cm}
\caption[]{Upper limits on the effective allowed halo fraction $f_{\rm
h,eff}$ as a result of taking into account the expectation fraction of
cluster stars which can be resolved $\langle f_{\rm res} \rangle$. The
assumed parameters are the same as for Fig.~\ref{f2}. Note that for
low cluster masses and large radii the allowed fraction is the same as
for Fig.~\protect{\ref{f2}}, whilst for large masses and small radii
$f_{\rm h,eff}$ becomes asymptotically large as $f_{\rm res}
\rightarrow 0$.}
\label{f3}
\end{figure}

Figure~\ref{f3} shows the upper limits on the effective halo fraction
$f_{\rm h,eff} \equiv 3\, f_{\rm c} N_{\rm c,eff}^{-1}$, assuming
$f_{\rm c} = 1$. Unlike the plot in Fig.~\ref{f2}, the surface does
not strictly represent a 95\%~CL constraint, since for that one would
need to evaluate the joint 95\%~CL variation in the cluster number
$N_{\rm c}$ and in the resolvable fraction $f_{\rm res}$. Such an
evaluation would require detailed Monte-Carlo simulations of the
cluster distribution for each parameter set $(m,M,R)$, and for each
field $i$.

Figure~\ref{f3} shows that the problem of cluster resolvability only
becomes important for high-mass, compact clusters, but that it can
have a very strong effect on detection in this regime. For the highest
mass and most compact clusters considered the resolvable fraction
$\langle f_{\rm res} \rangle$ rapidly approaches zero, so that no
point-source observation of a VLM star is expected, even if clusters
appear within the field of view. The density enhancement
$\epsilon_{\sigma}$ due to the superposition of clusters is found to
have little effect on $f_{\rm h}$, since $\epsilon_{\sigma}$ only
becomes significantly larger than unity for clusters with large radius
(in which case $\epsilon_{\sigma} \rightarrow \gamma$ in
Eq.~\ref{e13.2}), and these clusters anyway have such low surface
densities that the enhancement has little consequence.

As for Fig.~\ref{f2}, one can crudely parameterise the portion of the
plot which permits $f_{\rm h,eff}$ to be unity by using Eq.~\ref{e12}
with $\alpha \simeq 1.0$~pc and $\beta \simeq 0.73$ for 0.2-$\sm$
stars, or $\alpha \simeq 1.7$~pc and $\beta \simeq 0.69$ for
0.092-$\sm$ stars.  If one instead simply demands consistency with
MACHO ($f_{\rm h,eff} = 0.4$) then the allowed range is slightly
larger with $\alpha \simeq 1.2$~pc and $\beta \simeq 0.74$, or $\alpha
\simeq 2.5$~pc and $\beta \simeq 0.70$ for 0.2-$\sm$ or 0.092-$\sm$
stars, respectively.

\subsection{Dynamical considerations} \label{s3.4}

One can provide strong constraints on the permissible mass and radius
of clusters which contribute significantly to halo DM by
considering their dynamical effects on the visible stellar population
(e.g. Lacey \& Ostriker \cite{lac85}; Carr \& Lacey \cite{carr87};
Moore \cite{moo93}; Moore \& Silk \cite{moo95}). Such considerations
lead to the conclusion that viable cluster parameters $(M,R)$ are
bounded above and below, forming an `island' in parameter space
(Kerins \& Carr \cite{ker94}; Moore \& Silk \cite{moo95}). This
is important because it means that one cannot invoke arbitrarily
massive and compact clusters in order to force agreement between
various observations. Furthermore, in principle cluster parameters
which are compatible with dynamical limits may be incompatible with
source-count and MACHO limits, thereby ruling out the entire scenario.

The issue of dynamical constraints is somewhat complicated by the fact
that the limits depend (sensitively in some instances) on Galactic as
well as cluster parameters. Whilst I do not explicitly present the
dependencies of the constraints on these other parameters (these can
be found in the references cited above), I assume the following values
for the relevant parameters: a typical Galactocentric distance for
clusters in the vicinity of the HST fields of 8~kpc (in fact most of
the fields are further from the Galactic centre, where their dynamical
effects would be weaker than assumed here); a cluster halo
fraction $f_{\rm c} f_{\rm h,eff} = 0.4$ (required to provide
consistency with MACHO observations); an average halo mass density
$\rho_0 = 0.01~\sm~\mbox{pc}^{-3}$, giving a cluster number density
$f_{\rm c}f_{\rm h,eff} \rho_0/M$; a halo core radius $a = 5$~kpc; a
one-dimensional halo velocity dispersion of 156~km~s$^{-1}$; and a
Galaxy age of 15~Gyr. The dependency on Galactic parameters means that
the dynamical limits presented here do not represent firm limits in
the strictest sense, but should nonetheless help in assessing the
viability of the cluster scenario.  (Note that some of the authors
above express constraints in terms of the cluster half-mass radius
$R_{1/2} \simeq 0.3\, R$, assuming the surface-density profile of
Eq.~\ref{e14} with $r_{\rm c} = 0.1\, R$.)

An upper limit on the radius of clusters comes from the requirement
that they do not disrupt one another due to collisions occurring within
the lifetime of the Galaxy. This implies $R \la 100$~pc for the
parameters adopted above. Clusters also need to avoid tidal disruption
due to the differential force from the Galactic potential acting
across the cluster diameter; a constraint which requires $R \la 40\,
(M/10^4~\sm)^{1/3}$~pc. Since at any time there is always a finite 
fraction of cluster members whose velocities exceed the cluster escape
velocity, clusters slowly evaporate over time. That they do not
evaporate by the present day requires $R \ga 0.92~\mbox{pc}\,
(M/10^4~\sm)^{-1/3} [(m/0.2~\sm) \ln (0.4\,M/m)]^{2/3}$.

The final constraint comes from considering the effect of close or
direct collisions between dark clusters and observed diffuse globular
clusters (Carr \cite{carr78}; Moore \cite{moo93}; Moore \& Silk
\cite{moo95}). Moore (\cite{moo93}) showed that the requirement that
observed diffuse clusters show no sign of disruption today leads to
the limit $M \la 10^3~\sm$ for dark clusters with small radius, though
this value is derived from a rather unusually diffuse cluster that may
have atypical properties which result in too strong a constraint. A
more robust limit from other less diffuse and more representative
clusters demands $M \la 5 \times 10^4~\sm$ for $f_{\rm c}f_{\rm h,eff}
= 0.4$ (B.~Moore, private communication). For clusters with larger
radius Moore \& Silk (\cite{moo95}) derive a limit corresponding to $R
\ga 9\, (M/10^4~\sm)^{1/2}$~pc.

\begin{figure}
\epsfxsize=8.5cm
\epsfbox{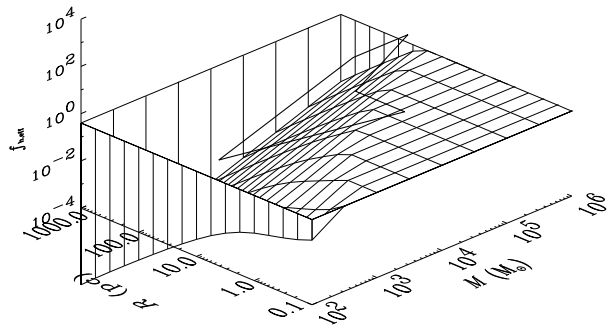}
%\picplace{8.0cm}
\caption[]{Comparison of HST limits, MACHO observations and dynamical
constraints, assuming $f_{\rm c} = 1$ and a VLM star mass of
$0.2~\sm$. The upper limits permissible from HST counts (shown in
Fig.~\protect{\ref{f3}}) for which $f_{\rm h,eff} > 0.4$ are projected
onto the plane $f_{\rm h,eff} = 0.4$; the central value inferred from
MACHO LMC observations for a local halo normalisation of $0.01~\sm
~\mbox{pc}^{-3}$. Also projected onto the same plane are the
dynamical limits discussed in the main text (bold lines), which are
sensitive to Galactic as well as cluster parameters. Note that there
exists a small region of parameter space with $M \sim 4 \times
10^{4}~\sm$ and $R \sim 3$~pc which is simultaneously compatible with
all of the limits. The overlap is marginally larger for lower
mass VLM stars.}
\label{f4}
\end{figure}

These limits, for an assumed star mass $m = 0.2~\sm$, are projected onto
the plane $f_{\rm h,eff} = 0.4$ in Fig.~\ref{f4} (bold lines),
together with the corresponding HST limits from Fig.~\ref{f3}, which
assume a clustering fraction $f_{\rm c} = 1$. Regions of parameter
space for which HST counts permit $f_{\rm h,eff} > 0.4$ are projected
onto the plane $f_{\rm h,eff} = 0.4$, corresponding to the halo
fraction preferred by MACHO microlensing observations towards the LMC
(Alcock et al. \cite{alc97}). Thus the flat region of the $(M,R)$
plane at small cluster radii and large masses simultaneously satisfies
MACHO observations and HST counts.  It is clear that this region and
the region bounded by the dynamical limits are almost mutually
exclusive, with only a small area of the $(M,R)$ plane satisfying all
three requirements. This is due to the strong limits inferred from the
absence of dynamical effects on presently-observed diffuse globular
clusters, and from evaporation considerations. The surviving parameter
region is characterised by a cluster mass $M \sim 4 \times 10^4~\sm$
and radius $R \sim 3$~pc and is relatively insensitive to the VLM star
mass within the range considered in this paper. Clearly, for the
cluster scenario to be viable $M$ and $R$ have to be rather finely
tuned, though it is interesting that the allowed parameter space
coincides well with the baryon Jeans mass at the recombination epoch,
as well as to the predictions of some baryonic DM formation theories
(e.g. Ashman
\cite{ash90}).

\subsection{Microlensing signatures} \label{s3.5}

The effect of clustering halo objects can be to increase or decrease
the observed microlensing rate, depending on whether or not the
observer's line of sight intersects with more or less than the
expected average number of clusters. One can therefore derive firm
upper and lower limits on the halo lens fraction by assuming it to
be clustered.

Maoz (\cite{maoz94}) has shown that there are two key features that
should help to discriminate between microlensing events in the
clustered and unclustered cases. The first is that the microlensing
optical depth through a cluster is $\tau \sim M R_{\rm e}^{2}/m
R^{2}$, where $R_{\rm e} \equiv [4 G m x(L-x)/c^2L]^{1/2}$ is the
Einstein radius and $L$ is the distance to the lensed source. Since
this can be a factor $10^{4}$ higher than in the unclustered case for
very massive ($M \sim 10^{6}~\sm$) compact ($R \sim 1$~pc) clusters,
Maoz notes that one expects to observe angular correlations between
events on a scale $\theta \sim R/x$. Maoz claims that for clusters of
this mass a sample as small as 10 events would be sufficient to rule
out such clusters from providing all the halo DM at the 95\%~CL, if
such correlations are not observed. Assuming that angular correlations
are observed, one should also expect nearby events to have similar
timescales, assuming the lenses have similar masses. This is because
the events would be due to objects residing in the same cluster, so
they would be at the same distance from the observer and would have
similar velocities. These conclusions are however based on cluster
mass scales which are now ruled out by the dynamical considerations
discussed in the previous subsection. The prospects for detecting such
signatures from clusters with $M \la 10^4 ~\sm$ are much less
optimistic (Metcalf \& Silk \cite{met96}), especially if they comprise
only 40\% of the DM.

Whether clustering could significantly affect the statistics for the
halo DM fraction $f_{\rm h}$ depends upon how many clusters one
expects to see within the solid angle of observation.  The MACHO LMC
search covers a solid angle of 11~deg$^{2}$ (Alcock et
al. \cite{alc96}), so the expected number of clusters between the
observer and the LMC ($l = 280\degr$, $b = -33\degr$, $L = 50$~kpc)
is
   \be
      N_{\rm exp} \simeq 1.3 \times 10^4 \, f_{\rm c}f_{\rm h}
      \left( \frac{M}{10^4~\sm} \right)^{-1} \!\!\! \left( \frac{\Omega}{
      11~\mbox{deg}^{2} } \right) \!\! \left(
      \frac{{\cal J}}{30} \right)\!, \label{e20}
   \ee
where
   \be
      {\cal J} \equiv \frac{1}{\rho_0 a^3} \int_0^L x^2 \rho(x,l,b) \,
      {\rm d}x \label{e21}
   \ee
for a halo with a density profile given by Eq.~\ref{e1}.  Thus, for
clusters which are simultaneously compatible with dynamical limits
and HST source counts (implying $M \sim 4 \times 10^4~\sm$
from Fig.~\ref{f4}) one expects $N_{\rm exp} \ga 1\,300$ to be within
the MACHO field of view if $f_{\rm c} = 1$ and $f_{\rm h} = 0.4$.

If the clusters all have the same mass and comprise objects with
the same MF, then the observed rate will depend only on
$N_{\rm exp}$. Assuming the cluster distribution on the sky obeys Poisson
statistics, the probability of there being $N$ clusters within the
field of view when one expects $N_{\rm exp}$ (in which case one would
infer a halo fraction $f_{\rm h} N/N_{\rm exp}$ when the true fraction is
$f_{\rm h}$) is simply $P(N,N_{\rm exp}) = (N_{\rm exp}^{N}/N!) \exp
(-N_{\rm exp})$.

The MACHO data yields a likely halo fraction $f_{\rm h} \simeq 0.4$ if
the lenses are assumed to be unclustered. The inferred value for
$f_{\rm h}$ would be the same for the cluster scenario provided $N =
N_{\rm exp}$, otherwise one should take $f_{\rm h} \simeq 0.4 \, N_{\rm
exp}/N$. The 1-$\sigma$ variation on $N_{\rm exp}/N$ is less than 3\%
for $N_{\rm exp} = 1\,300$, so the inferred halo fractions for the
clustered and unclustered regimes are virtually identical in this
case. Only for $f_{\rm c} \la 0.1$ do Poisson fluctuations in the
number of clusters become important, though in this case the
microlensing signal is anyway dominated by the smoothly distributed
VLM stars. The conclusion therefore is that for cluster mass scales
compatible with dynamical limits their microlensing signatures are
indistinguishable from the unclustered case.

It is interesting to consider what the inferred halo fraction from
microlensing, together with the HST source-count limits, imply for the
likely value for the clustering efficiency $f_{\rm c}$. The halo
considered in this paper has a microlensing optical depth $\tau = 5.6
\times 10^{-7}$ towards the LMC, assuming it is completely comprised
of lenses. The 95\%~CL lower limit on the measured optical depth is
$\tau_{95} \simeq 1.5 \times 10^{-7}$ (Alcock et al.  \cite{alc97}),
compared to the optical depth contribution expected from all non-halo
components of $\tau \la 5 \times 10^{-8}$. Subtracting the
contribution from these components gives a lower limit on the halo
fraction of $f_{\rm h} > (1.5-0.5)/5.6 = 0.17$. From Sect.~\ref{s2} the
maximum contribution from unclustered zero-metallicity VLM stars to
the halo is 1.4\% at the 95\%~CL. Therefore, consistency between MACHO
and HST observations requires that $f_{\rm c} > 1-(0.014/0.17) =
0.92$. That is, one requires a present-day clustering efficiency of
92\% or better. This is certainly a very strong demand for the cluster
scenario to meet.  If one adopts the central value for the measured
optical depth ($\tau = 2.9 \times 10^{-7}$) then the required
clustering efficiency must be at least 97\%.

\section{Discussion and conclusions} \label{s4}

In this paper I have considered the constraints on the contribution to
the halo dark matter (DM) of a population of zero-metallicity,
hydrogen-burning stars with mass below $0.2~\sm$ (VLM stars).

Though there are already strong constraints on the VLM star scenario,
this present work is motivated by 2 considerations: (1) previous
studies have adopted $V-I$ colour cuts for their star samples which
effectively eliminate any chance they have of discovering
zero-metallicity VLM stars; (2) previous studies only consider limits
on a smoothly distributed population of VLM stars, where as some
theories of baryonic DM formation predict that the stars should be
grouped into globular-cluster configurations.

Using data from 20 of the 22 HST fields obtained by Gould et
al. (\cite{gou96}), together with the photometric predictions of
Saumon et al. (\cite{sau94}) for zero-metallicity VLM stars, I find
that the contribution from a smoothly distributed population of such
stars to the halo can be no more than 1.4\% at the 95\% confidence
level on the basis of 75 candidate VLM stars with $1.2 < V-I < 1.7$
appearing in the fields when at least 6\,310 are predicted. In
reality, the true fraction is likely to be less than this value since
many of the candidates may belong to the spheroid or disc components.
This limit is comparable to previous analyses for stars of non-zero
metallicity and therefore the inescapable conclusion is that any
smoothly distributed population of VLM stars, regardless of its
metallicity, makes at best only a tiny contribution to the halo DM.

Clustering allows the possibility of much larger fluctuations in
the expected number of stars appearing in the fields. Additionally,
highly compact clusters may not be completely resolvable, thereby
decreasing the number of available point sources. These two effects
can permit a halo fraction in clusters which is compatible with
the halo fraction inferred by the MACHO gravitational microlensing
experiment provided that the cluster mass $M$ and radius $R$ satisfy
the inequality $R \la 1.2\, (M/10^4~\sm)^{0.74}$~pc.

One also requires that the cluster scenario satisfies the dynamical
constraints which exist on the allowed mass and radius of clusters.
Comparison of the dynamically allowed region with the region which
satisfies both HST and MACHO observations reveals that only a very
small portion of parameter space, characterised by $M \sim 4 \times
10^4~\sm$ and $R \sim 3$~pc, satisfies all three
requirements. However, the implicated cluster mass is intriguingly
close to the mass predicted by some DM formation theories (Ashman
\cite{ash90}), and to the baryon Jeans mass at the cosmological epoch
of recombination.

For a cluster mass $M \la 10^4~\sm$, which is required by dynamical
arguments, the effect on microlensing statistics is indistinguishable
from the unclustered case, and in particular the inferred halo
fraction should be within a few percent of the value inferred by
assuming the DM distribution to be unclustered. However, MACHO
and HST observations place a strong limit on the efficiency with which
VLM stars need to be clustered in order to remain compatible with
both surveys. This limit corresponds to a 92\% present-day clustering
efficiency and therefore provides a very stern test of the scenario.

Whilst recent microlensing results seemingly provide strong evidence
for a substantial baryonic contribution to Galactic halo DM, when
taken together with other observational and theoretical constraints it
has become increasingly difficult to provide a unique baryonic
candidate which can simultaneously explain the high microlensing
fraction and the event timescales.  One way out is to invoke a
substantial modification in the shape of the halo though, for this to
work, the microlensing results require a component resembling
something closer to a maximal disc, leaving the status of any halo
(and the role of non-baryonic DM on Galactic scales) much
reduced. Another option is to attribute the lensing events to some
non-baryonic candidate, such as primordial black holes, though in this
case one requires an additional non-baryonic candidate to explain the
rest of the halo DM. Here I have shown that it is still possible, if
only barely, to construct a baryonic scenario which is compatible with
all known constraints and which does not require a major modification
in the halo dynamics. It is also conceivable that the scenario could
provide an explanation for the recent detections of faint extended
emission around the edge-on spiral galaxy NGC~5907 (Sackett et
al. \cite{sac94}; Lequeux at al. \cite{leq96}) which, under the
assumption of constant mass-to-light ratio, appears to trace the
distribution of a halo.

\begin{acknowledgements}
I am grateful to Andy Gould, John Bahcall and Chris Flynn for their
permission to use the HST data, some of which is unpublished. I
particularly wish to thank Andy Gould for useful discussions and for
supplying the HST data in electronic form. I am also grateful to
Didier Saumon for providing the zero-metallicity VLM star model
predictions in electronic form, to Ben Moore for useful
correspondences, and to David Valls-Gabaud for helpful
discussions. This research is supported by a EC Marie Curie Training
\& Mobility of Researchers Fellowship.
\end{acknowledgements}

\end{document}